\documentclass[12pt]{iopart}

\usepackage{iopams}  

\usepackage{cite}
\usepackage{graphicx}
\usepackage{multirow}

\begin{document}

\title{One-parameter extension of the Doi-Peliti formalism and relation with orthogonal polynomials}

\author{Jun Ohkubo}

\address{
Graduate School of Informatics, Kyoto University,\\
36-1, Yoshida Hon-machi, Sakyo-ku, Kyoto-shi, Kyoto 606-8501, Japan
}
\ead{ohkubo@i.kyoto-u.ac.jp}
\begin{abstract}
An extension of the Doi-Peliti formalism for stochastic chemical kinetics is proposed.
Using the extension,
path-integral expressions consistent with previous studies are obtained.
In addition, the extended formalism is naturally connected to orthogonal polynomials.
We show that two different orthogonal polynomials, i.e.,
Charlier polynomials and Hermite polynomials,
can be used to express the Doi-Peliti formalism explicitly.
\end{abstract}

\pacs{05.40.-a, 82.20.-w, 02.50.Ey}
\maketitle

\section{Introduction}
\label{section_introduction}

Stochastic chemical kinetics has been widely used in various research areas.
For example, reaction-diffusion systems based on the stochastic chemical kinetics
have been used in nonequilibrium physics \cite{Tauber2005}.
In addition, recent development of experimental techniques enables us 
to observe various reactions in a cell in detail.
It has been clarified that
some chemical reactions in cells should be treated as systems with discrete states,
and continuous approximation cannot be used \cite{Rao2002}.
Hence, from the theoretical viewpoint,
it is important to develop tractable analytical schemes
to deal with the stochastic chemical kinetics.
The analytical schemes will give deep understanding
for the nonequilibrium phenomena,
and it is expected that
such analytical schemes lead to rapid computational method for parameter estimations,
which is useful for the analysis of experimental data.

The Doi-Peliti formalism is known as a useful method
to treat the stochastic chemical kinetics \cite{Doi1976,Doi1976a,Peliti1985}.
In the formalism, bosonic creation and annihilation operators
are used, and we can employ an analogy between the Doi-Peliti formalism
and quantum mechanics.
Actually, there are many researches for the Doi-Peliti formalism,
in which various techniques developed in quantum mechanics are used
in order to investigate classical stochastic process;
we can use perturbation calculations \cite{Dickman2003}, 
renormalization group analysis \cite{Tauber2005},
and system-size expansion \cite{Itakura2010}.

Although the usefulness of the Doi-Peliti formalism has been shown in many studies,
there are some unclear points of the formalism.
For example, is it possible to write state vectors, which are used in the Doi-Peliti formalism,
in terms of analytical functions?
If so, how should we interpret creation and annihilation operators?
While it has been pointed out that
the Doi-Peliti formalism is equivalent to generating function approach
or Poisson representation \cite{Droz1994},
the correspondence would not be unique.
If we have various concrete expressions for the Doi-Peliti formalism,
it is possible to choose an adequate one depending on one's objectives;
an expression may be suitable for perturbation calculations,
and another may be tractable for numerical calculations.

In the present paper,
we firstly propose an extension of the Doi-Peliti formalism.
While the extension includes an additional parameter,
the introduction of the parameter does not change
the coherent-state path-integral formula;
all analytical techniques in previous works are available.
The extension affect only the concrete expressions for state vectors in the formalism.
Secondly, we point out that
the one-parameter extension of the Doi-Peliti formalism
is naturally connected to orthogonal polynomials.
We find that two different orthogonal polynomials,
the Hermite polynomials and the Charlier polynomials,
are used to describe the state vectors in the formalism.

This paper is organized as follows.
In section~\ref{section_review},
a brief review of the Doi-Peliti formalism is given.
The correspondence between the Doi-Peliti formalism and the generating function approach
is also explained.
In section~\ref{section_extension}, we give the first main result
in the present paper, i.e., the one-parameter extension.
Section~\ref{section_orthogonal} is the second main result;
the Hermite polynomials and the Charlier polynomials are explained,
and concrete expressions of creation and annihilation operators are given.
Section~\ref{section_discussion} gives discussions and concluding remarks.

\section{A brief review of previous studies}
\label{section_review}

\subsection{The Doi-Peliti formalism}
\label{subsection_Doi-Peliti}

In the Doi-Peliti formalism, the following bosonic creation operator $a^\dagger$ 
and annihilation operators $a$ are used:
\begin{eqnarray}
[a,a^\dagger] \equiv a a^\dagger - a^\dagger a = 1, \quad [a, a] = [a^\dagger,a^\dagger] = 0, \label{eq_commutation},
\end{eqnarray}
where $[\cdot, \cdot]$ is the commutator,
and the actions of the creation and annihilation operators for ket vectors $| n \rangle$
are defined as
\begin{eqnarray}
a^\dagger | n \rangle = | n+1 \rangle, \quad a | n \rangle = n | n-1 \rangle. \label{eq_old_ket_vectors}
\end{eqnarray}
Here, the vacuum state $|0\rangle$ is characterized by $a | 0 \rangle = 0$.
\eref{eq_commutation} means that the creation operator
and annihilation operator do not commute.
While the actions of two operators on the ket vectors are defined 
as \eref{eq_old_ket_vectors},
actions on bra vectors $\langle n |$ are defined as follows:
\begin{eqnarray}
\langle n | a = \langle n+1 |, \quad \langle n | a^\dagger = \langle n-1 | n.
\label{eq_old_bra_vectors}
\end{eqnarray}
The inner product for the bra and ket vectors is given as
\begin{eqnarray}
\langle m | n \rangle = n! \, \delta_{m,n}.
\label{eq_old_bracket}
\end{eqnarray}
where $\delta_{m,n}$ is the Kronecker delta.

In order to explain the usefulness of the Doi-Peliti formalism,
we use a simple birth-coagulation process as an example.
The birth reaction $X \to X + X$ occurs with the rate constant $\alpha$
for each particle, and its backward reaction (coagulation) $X+X \to X$ occurs with $\beta$.
The master equation for the process is written as
\begin{eqnarray}
\frac{\partial}{\partial t} P(n,t) =& \alpha [(n-1)P(n-1,t) - n P(n,t)] \nonumber \\
&+ \beta [n (n+1) P(n+1,t) - n(n-1) P(n,t)],
\label{eq_master}
\end{eqnarray}
where $P(n,t)$ is a probability of finding $n$ particles at time $t$,
and we define $P(-1,t)\equiv 0$.
The remarkable idea of the Doi-Peliti formalism is 
the usage of a single vector $| \psi(t) \rangle$ which is a collection
of a series of infinite number of $P(n,t)$:
\begin{eqnarray}
| \psi(t) \rangle = \sum_{n=0}^{\infty} P(n, t) | n \rangle,
\end{eqnarray}
Using the vector $| \psi(t) \rangle$,
the master equation \eref{eq_master} is rewritten in a compact form
\begin{eqnarray}
\frac{\partial}{\partial t} | \psi(t) \rangle 
= L(a^\dagger,a) | \psi(t) \rangle,
\label{eq_Schrodinger_like}
\end{eqnarray}
where 
\begin{eqnarray}
L(a^\dagger,a) =
\alpha (a^\dagger - 1) a^\dagger a
+ \beta (1 - a^\dagger) 
a^\dagger a^2.
\label{eq_operator_example}
\end{eqnarray}
Because of the similarity with quantum mechanics,
the Doi-Peliti formalism is also called 
the second-quantization method or the field-theoretic approach.
In addition, similarities between the Doi-Peliti formalism and quantum mechanics enable us
to use various analytical techniques in quantum mechanics
in order to investigate classical stochastic processes,
as explained in section \ref{section_introduction}.

Here, we briefly explain some basic definitions
for a coherent-state path-integral expression,
which is especially useful in the Doi-Peliti formalism.
In order to derive the path-integral expression,
coherent states and a decomposition of unity play essential roles.
The coherent states are defined as
\begin{eqnarray}
| z \rangle \equiv \rme^{z a^\dagger} | 0 \rangle
= \sum_{n=0}^\infty \frac{1}{n!} z^n | n \rangle, \label{eq_old_coherent_state_ket}\\
\langle z | \equiv \langle 0 | \rme^{z^* a}
= \sum_{n=0}^\infty \frac{1}{n!} \langle n | (z^*)^n \label{eq_old_coherent_state_bra}
\end{eqnarray}
where $z$ is a complex number, and $z^*$ is the complex conjugate of $z$.
Using the coherent states,
the decomposition of unity is obtained as follows:
\begin{eqnarray}
\mathbf{1} = \sum_{n=0}^\infty \frac{1}{n!} | n \rangle \langle n |
= \sum_{n=0}^\infty \sum_{m=0}^\infty \frac{1}{m!} 
| n \rangle \langle m | \delta_{n,m} 
= 
\int \frac{\rmd^2 z}{\pi} \, \rme^{- |z|^2}
|z\rangle \langle z| \, ,
\label{identity_eq_old}
\end{eqnarray}
where we used 
\begin{eqnarray}
\delta_{n,m} =  \int \frac{\rmd^2 z}{\pi n!}\, 
\rme^{- |z|^2} z^{*m} z^n\, ,
\end{eqnarray}
with the integration measure 
$\rmd^2 z = \rmd(\mathrm{Re}\, z) \rmd(\mathrm{Im}\, z)$. 
Using the decomposition of unity,
it is straightforward to obtain the path-integral expression;
for details, see Ref. \cite{Tauber2005}.

Note that there are some differences between the Doi-Peliti method and quantum mechanics.
One of them is the calculation scheme for expectation values of observables.
Different from quantum mechanics,
an expectation value of observables in the Doi-Peliti formalism is obtained
by using a projection state:
\begin{eqnarray}
\langle \mathcal{P} | \equiv  \langle 0 | \mathrm{e}^{a} = \sum_{n=0}^\infty \frac{1}{n!} \langle n |.
\label{eq_old_projection_state}
\end{eqnarray}
For example, 
the average of $n$ is given by
$\sum_{n=0}^\infty n P(n,t) = \langle \mathcal{P} | a^\dagger a | \psi(t) \rangle$.

\subsection{Correspondence with the generating function approach}
\label{subsection_generating_function}

While the Doi-Peliti formalism is one of the analytical methods to treat
stochastic chemical kinetics,
the generating function approach is a more well-known method and has been studies a lot \cite{Gardiner_book}.
For the previous example in section~\ref{subsection_Doi-Peliti},
the generating function $G(x,t)$ is defined as
\begin{eqnarray}
G(x,t) = \sum_{n=0}^\infty P(n,t) x^n,
\end{eqnarray}
and the time evolution equation for $G(x,t)$ is written as
\begin{eqnarray}
\frac{\mathrm{d}}{\mathrm{d} t} G(x,t) 
= \left[
\alpha (x - 1) x \frac{\rmd}{\rmd x} 
+ \beta (1 - x) x \frac{\rmd^2}{\rmd x^2} 
\right] G(x,t).
\label{eq_time_evol_for_generating_function}
\end{eqnarray}
The above partial differential equation can be used
instead of the simultaneous differential equations \eref{eq_master}.

One may consider that \eref{eq_time_evol_for_generating_function}
is similar to \eref{eq_Schrodinger_like} with \eref{eq_operator_example}.
In fact, as written in section~\ref{section_introduction},
it has been pointed out that there is a correspondence between 
the Doi-Peliti formalism and the generating function approach \cite{Droz1994}.
When we interpret the creation and annihilation operators as
\begin{eqnarray}
a^\dagger \equiv x, \quad a \equiv \frac{\mathrm{d}}{\mathrm{d} x},
\end{eqnarray}
the linear operator $L(a^\dagger,a)$ for the time-evolution of $| \psi(t) \rangle$
in the Doi-Peliti formalism is obtained
from that of $G(x,t)$ by replacing $x$ and $\frac{\mathrm{d}}{\mathrm{d} x}$ 
with $a^\dagger$ and $a$, respectively.
In addition, explicit expressions for the ket and bra vectors in the Doi-Peliti formalism
are obtained as follows:
\begin{eqnarray}
| n \rangle \equiv x^n, \quad
\langle m | \equiv \int \mathrm{d} x \, 
\delta(x) \left( \frac{\mathrm{d}}{\mathrm{d} x} \right)^m (\cdot).
\end{eqnarray}
This correspondence has been used to discuss duality relations
in stochastic processes \cite{Ohkubo2010}.

\section{One-parameter extension of the Doi-Peliti formalism}
\label{section_extension}

Starting from the same commutation relation \eref{eq_commutation} and 
the same definitions for the ket vectors \eref{eq_old_ket_vectors},
it is possible to add one parameter $\lambda$ to the formalism;
this extension is one of the main results in the present paper.
That is, we define the following inner product instead of \eref{eq_old_bracket}:
\begin{eqnarray}
\langle m | n \rangle = \lambda^n n! \,\delta_{m,n}.
\label{eq_new_bracket}
\end{eqnarray}
According to the replacement of \eref{eq_old_bracket} with \eref{eq_new_bracket},
the actions of the creation and annihilation operators on bra vectors $\langle n |$ 
change as follows:
\begin{eqnarray}
\langle n | a = \langle n+1 | \lambda^{-1}, \quad \langle n | a^\dagger = \langle n-1 | n \lambda.
\label{eq_new_bra_vectors}
\end{eqnarray}
The derivation of \eref{eq_new_bra_vectors} is as follows:
Firstly, we have $\langle n+1 | a^\dagger | n \rangle = \langle n+1 | n+1 \rangle = (n+1) \lambda^{n+1}$
because $a^\dagger | n \rangle = | n+1 \rangle$.
Secondly, we assume the action of the creation operator on the bra vector
as $\langle n+1|a^\dagger = \langle n | \alpha$, where $\alpha$ is a scalar value.
Then, $\langle n+1 | a^\dagger | n \rangle = \alpha \langle n | n \rangle = \alpha n! \lambda^{n}$,
and we have $\alpha = (n+1)\lambda$.
Hence, the second equality in \eref{eq_new_bra_vectors} is obtained.
Using the similar discussions, the first equality in \eref{eq_new_bra_vectors} is easily checked.

We here note that the path-integral expression must not be changed 
due to the above one-parameter extension
because the final expression of the path-integrals
consists of integrals only for parameters in the coherent states, i.e., $z$ and $z^*$;
the final expression does not depend on the definition of $| n \rangle$ and $\langle n |$.
Actually, there is no need to change the definitions
of the projection state and the coherent states.
Since the actions of the creation and annihilation operators for the bra vectors
are modified as \eref{eq_new_bra_vectors},
we have $| n \rangle = (a^\dagger)^n | 0 \rangle$ and $\langle n | = \langle 0 | (a \lambda)^n$.
Hence, from the same definitions with the usual Doi-Peliti method,
we obtain slightly different expressions for the projection state and the coherent states
when we write them explicitly using the bra vectors $\langle n |$ as follows:
\begin{eqnarray}
\langle \mathcal{P} | \equiv  \langle 0 | \mathrm{e}^{a} 
= \sum_{n=0}^\infty \frac{1}{\lambda^n}\frac{1}{n!} \langle n |, \label{eq_new_projection_state}\\
| z \rangle \equiv \rme^{z a^\dagger} | 0 \rangle
= \sum_{n=0}^\infty \frac{1}{n!} z^n | n \rangle, \label{eq_new_coherent_state_ket}\\
\langle z | \equiv \langle 0 | \rme^{z^* a}
= \sum_{n=0}^\infty \frac{1}{n!} \langle n | \left( \frac{z^*}{\lambda^n} \right)^n. \label{eq_new_coherent_state_bra}
\end{eqnarray}
The above discussions suggest that
there is no need to change the definitions of the projection states and coherent states.
In addition, the decomposition of unity is calculated as 
\begin{eqnarray}
\mathbf{1} = \sum_{n=0}^\infty \frac{1}{\lambda^n}\frac{1}{n!} | n \rangle \langle n |
= 
\int \frac{\rmd^2 z}{\pi} \, \rme^{- |z|^2}
|z\rangle \langle z|.
\label{identity_eq_new}
\end{eqnarray}
The unity in the extended Doi-Peliti method has the same expression with the usual Doi-Peliti method
in terms of the coherent states,
and therefore we obtain the same path-integral expressions
even in the one-parameter extension, as expected.

\section{Correspondence with orthogonal polynomials}
\label{section_orthogonal}

As explained in section~\ref{subsection_generating_function},
the operators and vectors in the Doi-Peliti formalism 
are explicitly rewritten in terms of polynomials and differential operators.
The expression is not unique,
and we here show two different expressions 
for the one-parameter extension of the Doi-Peliti formalism.

In the following discussions, we restrict the additional parameter $\lambda$
as a positive real variable, i.e., $\lambda > 0$, 
in order to see the connection to the orthogonal polynomials.

\subsection{Hermite polynomials}
\label{section_Hermite}
One of the expressions is obtained from the Hermite polynomials \cite{Chihara_book}.
The Hermite polynomials are defined as
\begin{eqnarray}
H_n(x) = (-1)^n \rme^{x^2} \frac{\rmd^n}{\rmd x^n} \rme^{-x^2},
\end{eqnarray}
where $n \in \mathbb{N}, x \in \mathbb{R}$.
Introducing a scaling variable $\lambda \in \mathbb{R}$, 
we define the following rescaled Hermite polynomials:
\begin{eqnarray}
\tilde{H}^{(\lambda)}_n(x) \equiv \sqrt{ \left( \frac{\lambda}{2} \right)^n} 
H_n\left( \frac{x}{\sqrt{2\lambda}}\right).
\end{eqnarray}
Using the property of the Hermite polynomials,
it is straightforward to verify the following three-term recurrence formula:
\begin{eqnarray}
\tilde{H}^{(\lambda)}_{n+1}(x) 
= x \tilde{H}^{(\lambda)}_n(x) - \lambda n \tilde{H}^{(\lambda)}_{n-1}(x).
\label{eq_Hermite_recurrence_formula}
\end{eqnarray}
In addition, the rescaled Hermite polynomials satisfy the following orthogonality relation:
\begin{eqnarray}
\int_{-\infty}^{+\infty} \tilde{H}^{(\lambda)}_n(x) \tilde{H}^{(\lambda)}_m(x) 
\mu^{(\lambda)}(x) \rmd x
= \lambda^n n! \delta_{n,m},
\label{eq_Hermite_orthogonality}
\end{eqnarray}
where 
\begin{eqnarray}
\mu^{(\lambda)}(x) = \frac{1}{\sqrt{2\pi \lambda}} e^{- x^2 / (2\lambda)}.
\end{eqnarray}

As one can easily see,
the orthogonality relation \eref{eq_Hermite_orthogonality}
corresponds to the inner product \eref{eq_new_bracket}  
in the one-parameter extension of the Doi-Peliti method.
Actually, if we define
\begin{eqnarray}
| n \rangle \equiv \tilde{H}^{(\lambda)}_n(x), \quad
\langle n | \equiv \int_{-\infty}^{\infty} \rmd x \mu^{(\lambda)}(x) \tilde{H}^{(\lambda)}_n(x),
\end{eqnarray}
\begin{eqnarray}
a^\dagger \equiv x - \lambda \frac{\rmd}{\rmd x}, \quad
a \equiv \frac{\rmd}{\rmd x}.
\end{eqnarray}
all properties in the one-parameter extension of the Doi-Peliti formalism are recovered.
For example, 
the action of the creation operators on the bra vector, 
$\langle n | a^\dagger = \langle n-1 | n \lambda$,
is verified by using the recurrence formula \eref{eq_Hermite_recurrence_formula}
and a partial integral.

\subsection{Charlier polynomials}
\label{section_Charlier}
Another expression is obtained from the  Charlier polynomials \cite{Chihara_book}.
The definition of the monic Charlier polynomials is
\begin{eqnarray}
C^{(\lambda)}_n(x) = \sum_{k=0}^{n} (-\lambda)^{n-k} x^{(k)} {n \choose k}, 
\end{eqnarray}
where $x^{(k)} = x (x-1) \cdots (x-k+1)$
and $n \in \mathbb{N}, x \in \mathbb{N}$.
Note that the variable $x$ is not a real value but a natural number, 
which is different from the Hermite polynomials.
The Charlier polynomials satisfy the recurrence formula
\begin{eqnarray}
C^{(\lambda)}_{n+1}(x) = (x-n-\lambda) C^{(\lambda)}_n(x) - \lambda n C^{(\lambda)}_{n-1}(x)
\label{eq_Charlier_recurrence}
\end{eqnarray}
and the orthogonality relation
\begin{eqnarray}
\sum_{x=0}^{\infty} C^{(\lambda)}_n(x) C^{(\lambda)}_m(x) \frac{\lambda^x}{x!}\rme^{-\lambda}
= \lambda^n n! \delta_{m,n}.
\label{eq_Charlier_orthogonality}
\end{eqnarray}

We here introduce the following definitions for the bra and ket vectors:
\begin{eqnarray}
| n \rangle \equiv C^{(\lambda)}_n (x), \quad
\langle n | \equiv \sum_{x=0}^\infty \frac{\lambda^x}{x!}\rme^{-\lambda}
C^{(\lambda)}_n(x), \label{eq_Charlier_vectors}
\end{eqnarray}
In order to recover the properties of the one-parameter extension of the Doi-Peliti formalism,
we define the creation and annihilation operators as
\begin{eqnarray}
a^\dagger f(x) \equiv x f(x-1) - \lambda f(x), \quad
a f(x) \equiv f(x+1) - f(x). \label{eq_Charlier_operators}
\end{eqnarray}
Some techniques in discrete mathematics is needed
to verify \eref{eq_new_bra_vectors} for the above definitions,
and the explanations are a little complicated.
We will give them in Appendix.

\section{Discussions and concluding remarks}
\label{section_discussion}

We have found a one-parameter extension of the Doi-Peliti formalism,
and the extended formalism is deeply related to orthogonal polynomials.
Although the correspondence with the generating function approach
has already been known previously,
essentially different expressions for the one-parameter extension of the Doi-Peliti formalism
have been obtained.

While the extension does not affect the coherent path-integral formulation,
the additional parameter will be useful for some cases, for the following reason.
The time-evolution operator $L(a^\dagger,a)$ 
for the state vector $| \psi(t) \rangle$ is determined from the master equation,
and it is independent from the additional parameter $\lambda$.
On the other hand, the parameter $\lambda$ can change analytical expressions
for the bra and ket vectors, and hence analytical expressions of the state vector $| \psi(t) \rangle$
depend on the parameter $\lambda$.
As a result, it is expected that
one can choose an adequate parameter $\lambda$ 
which is suitable to express the solution $|\psi(t)\rangle$.
Actually, it has been known that 
some stochastic processes are related to orthogonal polynomials,
and the choice of the parameters in the orthogonal polynomials
is important to express the analytical solutions \cite{Schoutens_book}.

In the correspondence between the generating function approach and the usual Doi-Peliti method,
the expression with the Dirac's delta function has been known.
However, the delta function may be intractable 
when we want to construct some numerical methods based on the formalism.
In contrast, the Hermite and Charlier polynomials would be tractable,
and hence they will become helpful in numerical computations.
As explained in section~\ref{section_introduction},
the Doi-Peliti formalism has been widely used.
We expect that the correspondence to the orthogonal polynomials will give useful methods
in order to investigate nonequilibrium behaviour.

\section*{Acknowledgments}
This work was supported in part by grant-in-aid for scientific research 
(Grant No.~20115009)
from the Ministry of Education, Culture, Sports, Science and Technology (MEXT), Japan.

\appendix
\section{Validity of the definitions in section \ref{section_Charlier}}

It is well-known that the difference operator \cite{Graham_book}
\begin{eqnarray}
\Delta u(x) \equiv u(x+1) - u(x)
\end{eqnarray}
acts on the Charlier polynomials as
\begin{eqnarray}
\Delta C_n^{(\lambda)}(x) = C_n^{(\lambda)}(x+1) - C_n^{(\lambda)} (x)
= n C_{n-1}^{(\lambda)} (x),
\label{app_annihilation}
\end{eqnarray}
so that $a | n \rangle = n | n-1 \rangle$ is verified.
In addition, combination of the recurrence formula \eref{eq_Charlier_recurrence}
and the difference equation \cite{Chihara_book}
\begin{eqnarray}
-n C_n^{(\lambda)}(x) = \lambda C_n^{(\lambda)}(x+1) - (x+a) C_n^{(\lambda)}(x)
+ x C_n^{(\lambda)}(x-1)
\end{eqnarray}
gives 
\begin{eqnarray}
C_{n+1}^{(\lambda)}(x) = x C_n^{(\lambda)}(x-1) - \lambda C_n^{(\lambda)}(x),
\end{eqnarray}
which corresponds to $a^\dagger | n \rangle = | n+1 \rangle$.

In order to check $\langle n | a^\dagger = \langle n-1 | n \lambda$,
a partial summation is available.
Using the shift operator defined as
\begin{eqnarray}
\mathrm{E} u(x) \equiv u(x+1),
\end{eqnarray}
the partial summation is given by \cite{Graham_book}
\begin{eqnarray}
\sum u \Delta v = u v - \sum \Delta u \mathrm{E}v .
\end{eqnarray}
We here note that we can interpret $a^\dagger | n \rangle$ as
\begin{eqnarray*}
x C_n^{(\lambda)}(x-1) - \lambda C_n^{(\lambda)}(x)
= - x \Delta C_n^{(\lambda)}(x-1) + x C_n^{(\lambda)}(x) - \lambda C_n^{(\lambda)}(x).
\end{eqnarray*}
Hence, $\langle n | a^\dagger | m \rangle$ is expressed as
\begin{eqnarray}
\fl
\sum_{x=0}^\infty \frac{\rme^{-\lambda} \lambda^x}{x!} C_n^{(\lambda)}(x) 
\left[ - x \Delta C_m^{(\lambda)}(x-1) + x C_m^{(\lambda)}(x) - \lambda C_m^{(\lambda)}(x) \right] \nonumber \\
\fl
= \left[
\frac{\rme^{-\lambda} \lambda^x}{x!}  C_n^{(\lambda)}(x)  (-x) C_m^{(\lambda)}(x-1) 
\right]_{x=0}^{x=\infty} \nonumber \\
\fl
\quad + \sum_{x=0}^\infty \left[
\Delta \left\{ \frac{\rme^{-\lambda} \lambda^x}{x!} C_n^{(\lambda)}(x)  x \right\}
\mathrm{E} C_m^{(\lambda)}(x-1)
+ \frac{\rme^{-\lambda} \lambda^x}{x!}  C_n^{(\lambda)}(x) \left\{
x C_m^{(\lambda)}(x) - \lambda C_m^{(\lambda)}(x)
\right\}
\right] \nonumber \\
\fl = 
\sum_{x=0}^\infty 
\left[ \left\{ \frac{\rme^{-\lambda} \lambda^{x+1}}{x!} C_n^{(\lambda)}(x+1)
-  \frac{\rme^{-\lambda} \lambda^x}{(x-1)!} C_n^{(\lambda)}(x)
\right\} C_m^{(\lambda)}(x)
\right. \nonumber \\
\fl \left. 
\quad + \frac{\rme^{-\lambda} \lambda^x}{x!}  C_n^{(\lambda)}(x) \left\{
x C_m^{(\lambda)}(x) - \lambda C_m^{(\lambda)}(x)
\right\}
\right] \nonumber \\
\fl = 
\sum_{x=0}^\infty 
\frac{\rme^{-\lambda} \lambda^x}{x!} \lambda 
\left\{ C_n^{(\lambda)}(x+1) - C_n^{(\lambda)}(x) \right\} C_m^{(\lambda)}(x) \nonumber \\
\fl = 
\sum_{x=0}^\infty 
\frac{\rme^{-\lambda} \lambda^x}{x!} \lambda n C_{n-1}^{(\lambda)}(x)  C_m^{(\lambda)}(x) ,
\end{eqnarray}
where we used \eref{app_annihilation} to obtain the final equality.
Hence, the action of the creation operator $a^\dagger$ 
on the bra vector, $\langle n | a^\dagger = \langle n-1 | n \lambda$,  is verified.
The action of the annihilation operator $a$ on the bra vector,
$\langle n | a = \langle n+1 | \lambda^{-1}$
can be checked in a similar manner.

\end{document}